# Retrieval-as-a-Service:
# A System-Oriented Analysis of Industrial Retrieval Pipelines in Web Systems

Fang Liu[1*], Yuan Yuan[2], Yifan Dang[3], Xuncheng Zhang[4], Cuiqianhe Du[5]

1 Yale University, USA
2 Stevens Institute of Technology, USA
3 University of Maryland College Park, USA
4 University of California, Berkeley, USA
5 University of California, Berkeley, USA

* Corresponding author: fangliu435@gmail.com

Retrieval systems have become a foundational infrastructure component in modern Web services, supporting applications such as content recommendation, advertising targeting, and API discovery. In large-scale industrial environments, retrieval is increasingly deployed as an independent service layer, commonly referred to as Retrieval-as-a-Service (RaaS).

This paper presents a system-oriented survey of industrial retrieval pipelines, focusing on architectural design and deployment trade-offs under real-world constraints. Unlike prior surveys that emphasize algorithmic developments, we analyze retrieval systems from an infrastructure perspective, highlighting how latency requirements, scalability constraints, and resource limitations shape system design in production environments.

We introduce a unified RaaS pipeline abstraction that models retrieval as a multi-stage service, including high-efficiency candidate generation, embedding-based semantic matching, and resource-aware re-ranking. We further examine the integration of Large Language Model (LLM)-based retrieval mechanisms and analyze their impact on semantic performance, latency, and computational overhead.

The results provide a system-level understanding of retrieval as a service-oriented infrastructure and offer practical guidelines for designing scalable, efficient, and QoS-aware retrieval architectures in large-scale Web systems.

# 1 Introduction

Delivering personalized content and advertisements is central to user engagement in modern Web systems [1–4]. In large-scale platforms, retrieval mechanisms are widely used to match user intent with relevant content under strict latency and scalability requirements. However, as these systems evolve, retrieval is no longer treated as a standalone algorithmic component, but increasingly as part of a distributed infrastructure that must operate under real-world constraints.

In contemporary Service-Oriented Architectures (SOA), retrieval functions as an independent microservice responsible for selecting relevant candidates from massive corpora in real time [4–6]. This paradigm, which we refer to as Retrieval-as-a-Service (RaaS), reflects a shift from model-centric design toward system-level optimization. In such environments, retrieval systems must satisfy stringent Service Level Agreements (SLAs) and Quality-of-Service (QoS) requirements, often within millisecond-level latency budgets [7, 8].

While extensive research has focused on improving retrieval algorithms and recommendation models, most existing studies emphasize algorithmic performance, such as ranking accuracy and representation learning [9–13]. In contrast, relatively fewer works examine how these methods are deployed and orchestrated in production environments, where system-level constraints such as throughput, memory footprint, and hardware availability significantly influence design decisions.

This paper builds upon prior work on retrieval algorithm categorization [9], but shifts the focus toward a service-oriented perspective. Rather than analyzing individual models in isolation, we investigate how different retrieval paradigms—such as index-based methods, embedding-based matching, and generative retrieval—are integrated into multi-stage industrial pipelines under real-world constraints.

The main contributions of this work are summarized as follows:

1. We propose a service-oriented abstraction for modeling industrial retrieval pipelines as Retrieval-as-a-Service (RaaS) systems.
2. We provide a systematic analysis of multi-stage retrieval architectures, highlighting the trade-offs between semantic performance and system efficiency under QoS constraints.
3. We examine the integration of emerging paradigms, including Large Language Model (LLM)-based retrieval, and discuss their implications for large-scale industrial deployment.

This survey is intended for researchers, practitioners, and system architects working on industrial information retrieval, recommendation systems, and large-scale Web services, particularly those involved in designing and deploying RaaS pipelines under real-world operational constraints. By emphasizing system-level considerations, this work complements existing algorithm-focused studies and provides a more comprehensive understanding of retrieval systems in practice.

## 2 Related Work

Research on retrieval for advertising and recommendation spans several decades, evolving from simple keyword matching to complex neural architectures. Early studies in advertising focused on behavioral signals [1] and index compression [10] to optimize relevance under millisecond-level constraints. Similarly, organic recommendation evolved from collaborative filtering and matrix factorization [11] , [12] toward deep Two-Tower embedding models [13] and multi-task learning frameworks [9].

Emerging trends such as Retrieval-Augmented Generation (RAG) [14-17] and LLM-driven pipelines further highlight the convergence of retrieval components across diverse intelligent systems. Beyond traditional content, these paradigms are increasingly applied to Web Service discovery and API recommendation, where retrieval functions as a critical matching service for functional endpoints. Unlike previous surveys [18-21] that treat these domains in isolation, this work provides a unified service-centric framework focusing on the trade-offs between algorithmic complexity and industrial Quality-of-Service (QoS) requirements.

Beyond content delivery, retrieval paradigms are increasingly applied to service-oriented tasks. For instance, recent work at ICWS 2024 explored biased random walks in heterogeneous networks for Web API recommendation [5], highlighting the convergence of retrieval and service discovery. Several recent surveys have explored neural retrieval and recommendation systems [20], [21], with more recent studies further extending these directions [37], [38]. However, most prior work primarily focuses on algorithmic improvements, while limited attention has been given to service-level deployment constraints and QoS trade-offs in large-scale industrial Web systems. This gap motivates the service-centric analysis presented in this paper. In this work, these developments are not analyzed as standalone algorithmic innovations, but are instead treated as components within industrial retrieval pipelines subject to system-level constraints.

## 3 Survey Methodology

To ensure comprehensive and unbiased coverage of the literature, we conducted a structured search across multiple academic databases, including Google Scholar, IEEE Xplore, ACM Digital Library, which together cover both peer-reviewed publications and widely recognized preprint repositories.

The search was performed using combinations of keywords such as “information retrieval systems,” “industrial retrieval pipelines,” “ad targeting systems,” “recommendation systems,” and “retrieval architectures in web services.” The review primarily focuses on publications from 2015 to 2025 to capture recent advances in large-scale industrial retrieval systems, while also incorporating earlier foundational works where necessary to provide historical context. The selection process was iterative, with references of identified studies further examined to locate additional relevant work.

The inclusion criteria were: (1) studies describing large-scale retrieval or recommendation systems, (2) works providing architectural or algorithmic insights relevant to industrial deployment, and (3) publications from peer-reviewed venues or widely recognized preprint repositories. Studies that were purely theoretical without practical deployment context or not directly related to retrieval pipelines were excluded.

To reduce selection bias, we cross-validated findings across multiple sources and prioritized highly cited and recent works. In addition, both academic research and industry-driven reports were considered to provide a balanced perspective on retrieval system design and deployment.

This methodology ensures systematic and representative coverage of both theoretical developments and real-world industrial practices. The final selection includes a curated set of representative works covering both classical retrieval methods and recent advances in neural and LLM-based retrieval systems, as reflected in the references cited in this paper. While this study does not aim to be a fully exhaustive systematic review, it follows a structured and reproducible methodology to ensure representative and balanced coverage of both academic research and industrial practices. Unlike prior surveys that primarily focus on algorithmic advancements, this work emphasizes how these retrieval techniques are orchestrated within large-scale service architectures under real-world system constraints.

## 4 Ad Targeting Strategies

In this work, traditional targeting strategies are abstracted as modular components within industrial retrieval pipelines, rather than being analyzed purely from an algorithmic perspective. Ad targeting has evolved from simple rule-based matching into complex discovery services that leverage massive user signals. In a Service-Oriented Architecture (SOA), these strategies function as specialized retrieval filters or embedding-based matching services to fulfill advertising Service Level Agreements (SLAs) [11-13].

We categorize the predominant industrial targeting strategies in Table 1, mapping their business objectives to their underlying retrieval implementations. By viewing these as functional microservices, platforms can achieve high-concurrency matching while maintaining strict QoS.

Table 1. Mapping Industrial Targeting Strategies to Retrieval Implementations

| Targeting Strategy | Core Data Signals | Service-Level Implementation | Industrial Deployment Considerations |
|---|---|---|---|
| Demographic | Age, Gender, Language | Boolean Filtering / Inverted Index | Fast lookup, sub-millisecond latency, easy horizontal scaling |
| Contextual | Location, Device, Time | Real-time Geo-hashing Services | Low-latency, geo-distributed scaling required |
| Behavioral | Clicks, Conversion Events | Two-tower Embedding RaaS | GPU-intensive, SLA-sensitive, high memory usage |
| Retargeting | Past Interactions | Hash-based Lookup / ID Mapping | Simple to deploy, limited semantic matching |
| Keyword/LLM | Search Query, Intent | LLM-driven Intent Expansion | Heavy model, latency/throughput trade-offs, requires model serving infrastructure |

## 5 Retrieval Frameworks for Ad Targeting and Recommendation

Large-scale recommendation and advertising systems are often modeled as information retrieval problems, where users and items are represented in structured indices or learned embedding spaces [3], [10], [22]. In industrial Web Services environments, retrieval pipelines must balance semantic matching quality with latency, throughput, and memory constraints.

A typical RaaS pipeline is illustrated in Figure 1, where retrieval is organized as a multi-stage service consisting of feature-augmented candidate generation, embedding-based retrieval, approximate nearest neighbor (ANN) search, and multi-stage re-ranking. Each stage reflects a trade-off between efficiency and semantic expressiveness under industrial QoS constraints. This pipeline representation abstracts common industrial retrieval architectures and serves as a unifying framework for analyzing RaaS systems.

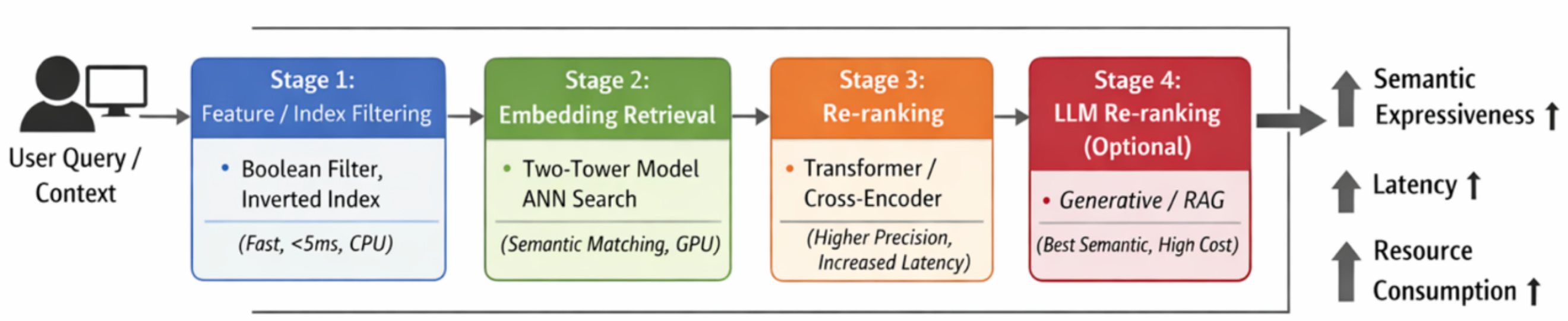

Figure 1. A typical multi-stage RaaS pipeline in industrial web systems

*5.1 Problem Formulation and Retrieval Overview*

In large-scale platforms, retrieval serves as the first-stage filtering mechanism that selects a small candidate set from a massive corpus of ads or items. Given a user context, including demographic attributes, historical behaviors, and contextual signals, the system selects candidates likely to be relevant under strict latency and scalability constraints. Formally, the retrieval service is defined as a mapping function:

$$R : (u, c) \rightarrow \{a^1, \dots, a^k\}$$

$$A^* = \text{argmax}_{\{a \in A\}\text{Score}(u,a)}$$

$$\text{Latency}(R) \leq \tau_{\{SLA\}}$$

where $\tau_{(SLA)}$ denotes the latency constraint defined by SLA requirements under QoS constraints in large-scale industrial systems.

*5.2 Representative Retrieval Methods*

Index-based retrieval methods, such as inverted indexes, form the foundation of many industrial ad targeting systems [10]. In these systems, items are organized by discrete attributes or

keywords, enabling efficient sub-millisecond lookup under strict SLA constraints.

Feature-augmented retrieval approaches extend this paradigm by incorporating richer item metadata and user signals. While these methods are computationally efficient and interpretable, they are typically used as initial "hard-filter" stages in modern RaaS architectures.

As illustrated in Table 2, these retrieval paradigms reflect a fundamental trade-off between semantic expressiveness and industrial QoS requirements.

Table 2. Industrial Roles and QoS Profiles of Retrieval Methods

| Method | Strengths | Limitations | Industrial Role & QoS Profile |
|---|---|---|---|
| **Inverted Index** | Fast lookup; highly scalable | Limited semantic matching | **Role:** High-speed filtering. **QoS:** < 5ms latency, CPU-bound. |
| **Two-Tower** | Scalable embedding matching | Limited cross-feature interaction | **Role:** Semantic discovery. **QoS:** GPU-accelerated, Memory-heavy. |
| **Transformer** | Strong representation | High computational cost | **Role:** Precise re-ranking. **QoS:** High latency, GPU intensive. |

Neural retrieval models, including the Two-Tower and Transformer-based architectures [25], improve representation capacity and personalization at the cost of increased training and system complexity. Modern industrial systems frequently employ a multi-stage pipeline to balance these competing objectives, using index-based methods for broad filtering and neural models for semantic discovery. Generative retrieval approaches, such as Large Language Models (LLMs) augmented with structured knowledge or industrial pipelines, can further enhance semantic relevance, though they introduce additional latency and GPU requirements [15], [23], [24].

*5.3 Training and Inference*

Training objectives in industrial retrieval systems typically include cross-entropy loss and pairwise ranking loss, which are designed to optimize relevance estimation and ranking quality [12], [25], [26]. In practice, these objectives are often combined with large-scale negative sampling strategies to improve generalization in high-dimensional embedding spaces.

During inference, retrieval systems must balance model complexity with strict industrial QoS constraints. To achieve this, embeddings are commonly pre-computed and stored in distributed index structures, enabling efficient lookup. Approximate nearest neighbor (ANN) search is widely adopted to ensure sub-millisecond retrieval latency [3], even when operating over billions of candidates in industrial-scale systems [4], [5]. These design choices reflect the need to decouple training complexity from real-time serving requirements in RaaS architectures.

*5.4 Challenges and Extensions*

Industrial retrieval systems face several persistent challenges in large-scale deployment.

First, cold-start scenarios remain a fundamental issue, as new users or items often lack sufficient interaction history, making it difficult for retrieval models to generate accurate representations.

Second, high-throughput and memory constraints pose significant challenges. Embedding-based and Transformer-based models require substantial computational resources, and data sparsity and noise can further degrade embedding quality in large-scale systems.

Third, incorporating multi-task learning and contextual signals introduces additional complexity. While shared representations across multiple objectives can improve generalization, they also increase system design complexity and training cost.

Recent industrial experiences indicate that LLM-based retrieval can improve semantic matching, but requires careful GPU allocation and latency management [23], [24], [28].

### *5.5 Industrial Deployment Trade-offs*

In production environments, each retrieval method presents distinct trade-offs across latency, scalability, and semantic capability. Index-based methods offer extremely low latency and efficient horizontal scaling, making them well suited for high-throughput filtering scenarios, although they are limited in capturing deep semantic relationships. Feature-augmented methods improve interpretability and leverage richer metadata signals, but remain constrained in modeling complex user intent and nuanced semantic interactions. Two-Tower embedding models provide strong semantic expressiveness and scalable matching capabilities; however, they are resource intensive and often sensitive to SLA constraints due to their reliance on GPU acceleration. Transformer-based and generative retrieval approaches further enhance precision and adaptability for complex queries, but introduce significantly higher latency and computational overhead, making them less suitable for real-time serving in latency-critical systems.

The comparison in Table 3 illustrates a fundamental trade-off among semantic expressiveness, latency, and resource efficiency.

Table 3. Industrial Comparison of Retrieval Methods

| Method | Latency | Throughput | GPU/CPU | Memory | Cold-start | Semantic |
|---|---|---|---|---|---|---|
| Index | <5ms | High | CPU | Low | Poor | Low |
| Two-Tower | 10–50 ms | Medium | GPU | Medium | Medium | High |
| Transformer | 50–200 ms | Low | GPU | High | Medium | Very High |
| Generative | 100–500 ms | Low | GPU | High | Medium | Very High |

In practical deployment scenarios, different configurations are preferred depending on system constraints. In hard-constraint environments with sub-5ms latency requirements (e.g., real-time bidding), index-based methods remain the only viable option due to their low-latency, CPU-bound nature. In contrast, semantic-first scenarios such as Web API recommendation often require Two-Tower or Transformer-based models to achieve higher intent understanding, despite increased computational overhead. Generative retrieval approaches further enhance adaptability, but are typically more suitable for asynchronous enrichment rather than real-time retrieval paths due to their significant latency and resource demands.

## 6 Comparing Ad Targeting and Organic Retrieval

While ad targeting and organic retrieval share technical foundations in large-scale indexing and embedding-based matching, they diverge significantly in their service objectives and operational constraints [1], [7]. Ad targeting serves as a revenue-driving engine, whereas organic retrieval focuses on ecosystem health and user retention. This section analyzes these paradigms through a unified retrieval-centric perspective.

These differences highlight the need for distinct retrieval strategies under different service objectives. To clarify these differences, Table 4 summarizes a systematic comparison between ad targeting and organic retrieval across key dimensions.

Table 4. Systematic Comparison: Ad Targeting vs. Organic Retrieval

| Dimension | Ad Targeting (Paid) | Organic Retrieval (Unpaid) |
|---|---|---|
| **Core Objective** | Revenue maximization, CTR, and conversion. | User satisfaction, diversity, and long-term engagement. |
| **Data Granularity** | Extensive demographic and contextual profiling. | Historical preferences and consumption patterns. |
| **Revenue Logic** | Direct monetization via RTB and auctions. | Indirect value through retention and loyalty. |
| **Constraint** | Ad fatigue, trust, and strict privacy ethics. | Data sparsity and cold-start discovery. |

The divergence in these dimensions necessitates different Service-Level Agreement (SLA) priorities. For ad targeting, the retrieval service must handle the complexity of real-time auction mechanisms within milliseconds. In contrast, organic retrieval services often prioritize the "explorative" nature of the results, requiring more complex models to handle the cold-start problem for new content. Understanding these differences is crucial for designing a unified RaaS platform that can support both paradigms efficiently.

## 7 Evaluation Metrics and Deployment Considerations

This section examines evaluation metrics and deployment trade-offs in industrial RaaS pipelines, emphasizing service-oriented QoS considerations rather than controlled experimental results.

Evaluating RaaS in production environments requires a paradigm shift from purely algorithmic metrics to service-oriented QoS indicators.

*7.1 Multi-Dimensional Evaluation Framework*

We define a holistic evaluation framework that segregates metrics by service intent (Organic vs. Paid) while unifying them under industrial constraints. The key dimensions of this framework are summarized in Table 5.

1. SLA & QoS Priority: In RaaS, the P99 latency and SLA compliance rate often outweigh minor gains in AUC. Our framework prioritizes "fixed-budget optimization," where models are selected based on their ability to stay within a 5-20 ms window.
2. Resource Footprint: Unlike academic benchmarks, we explicitly measure GPU/Memory utilization. This is critical for horizontal scaling in cloud-native environments where resource contention can degrade overall system throughput.

Table 5. Industrial Evaluation Metrics for RaaS Pipelines

| **Dimension** | **Organic Recommendation** | **Ad Targeting (Paid)** | **Industrial Deployment Considerations** | **LLM Integration Impact** |
|---|---|---|---|---|
| Primary Metrics | DAU/MAU, Retention, Interaction | Revenue, CPC, Conversion Rate | SLA compliance, latency distribution, throughput | Latency ↑5–20ms, Semantic relevance ↑10–15% |
| User Signal | Implicit interest, long-term engagement | Ad relevance vs. intrusiveness | Load balancing, memory footprint, GPU utilization | GPU usage ↑40–60%, small increase in memory footprint |
| Experimental Focus | Traffic interference, long-term engagement | Budget control, market dynamics | A/B testing with SLA monitoring | Can simulate LLM impact using subset of queries |
| System Constraints | Content discovery, quality | Advertiser ROI, bidding fairness | Cold-start handling, horizontal scaling, fault tolerance | Must consider SLA trade-offs and throughput reduction |

*7.2 Performance Trade-offs in LLM-Augmented Retrieval*

To provide practical insights into the deployment trade-offs of RaaS pipelines, we analyze the impact of integrating LLM-based reranking within a typical high-concurrency retrieval architecture. In such systems, a baseline Two-Tower retrieval model [4] is often combined with an LLM-based reranking component [23], [24], [28]. Based on observations from industrial retrieval systems and prior studies, this integration typically improves semantic relevance but introduces additional latency and GPU utilization overhead.

*7.2.1 Quantitative Analysis*

Prior studies demonstrate that deeper semantic modeling often comes with increased operational cost in large-scale retrieval systems.

1. Semantic vs. Latency: While LLM integration boosts semantic relevance by 10–15% (particularly for long-tail queries), it introduces a 5–20ms latency penalty. In a synchronous RaaS path, this overhead can trigger SLA violations if not mitigated by aggressive caching.
2. Infrastructure Stress: Prior studies and industrial observations suggest a potential 40–60% increase in GPU utilization. This "Resource Tax" suggests that for high-traffic Web services, LLM components should be deployed as asynchronous enrichment services to maintain system resilience.

### *7.2.2 Summary of Deployment Insights*

The primary takeaway from our observations is that the optimal RaaS configuration is SLA-aware. For real-time bidding (RTB) where budgets are <10ms, simpler feature-augmented models remain dominant, whereas API Recommendation services can better tolerate the LLM latency overhead to achieve higher intent satisfaction.

## 8 Conclusion

This work provides a unified service-oriented abstraction of industrial retrieval pipelines, positioning RaaS as a core architectural paradigm in modern Web systems. By moving beyond individual algorithms toward service-level architectures, industrial retrieval systems are fundamentally shaped by the need to balance semantic expressiveness with strict QoS requirements. As Web systems continue to scale, performance is increasingly constrained not only by model precision, but also by the operational limitations imposed by SLAs.

The analysis highlights several implications for both research and practice. Integrating LLM-based retrieval components, such as RAG and reranking modules, can improve semantic relevance, particularly for complex or long-tail queries. These improvements, however, come with additional latency (typically 5–20 ms) and increased GPU utilization, requiring careful trade-offs in production environments. Multi-stage retrieval pipelines therefore remain a practical necessity, enabling systems to balance efficiency and effectiveness across different stages. More adaptive orchestration strategies—where retrieval pathways are adjusted based on real-time system conditions—may offer a promising direction. At the same time, growing concerns around data privacy and resource constraints point toward the need for more efficient and privacy-preserving retrieval architectures, including federated and on-device approaches.

The transition toward unified RaaS architectures represents an important step in the evolution of large-scale Web systems. Future work may further explore adaptive pipeline design, resource-aware deployment of LLM-based retrieval components, and scalable architectures capable of operating reliably under real-world production constraints.